\documentclass[10pt,journal]{IEEEtranTCOM}

\usepackage[english]{babel}
\usepackage[usenames]{color}
\usepackage[cp1250]{inputenc}
\usepackage{amsfonts}
\usepackage{amsthm}
\usepackage{graphicx}
\usepackage{epsfig}
\usepackage{mathrsfs}
\usepackage{amsmath}
\usepackage{algorithm}
\usepackage{algorithmic}
\usepackage{hyperref}

\theoremstyle{plain}

\begin{document}
\newcommand{\bea}{\begin{eqnarray}}
\newcommand{\eea}{\end{eqnarray}}
\newcommand{\be}{\begin{equation}}
\newcommand{\ee}{\end{equation}}
\newcommand{\beas}{\begin{eqnarray*}}
\newcommand{\eeas}{\end{eqnarray*}}
\newcommand{\bs}{\backslash}
\newcommand{\bc}{\begin{center}}
\newcommand{\ec}{\end{center}}
\def\SC {\mathscr{C}}

\title{Predicting probability distributions for\\ cancer therapy drug selection optimization}
\author{\IEEEauthorblockN{Jarek Duda}\\
\IEEEauthorblockA{Jagiellonian University,
Golebia 24, 31-007 Krakow, Poland,
Email: \emph{dudajar@gmail.com}}}
\maketitle

\begin{abstract}
Large variability between cell lines brings a difficult optimization problem of drug selection for cancer therapy. Standard approaches use prediction of value for this purpose, corresponding e.g. to expected value of their distribution. This article shows superiority of working on, predicting the entire probability distributions - proposing basic tools for this purpose. We are mostly interested in the best drug in their batch to be tested - proper optimization of their selection for extreme statistics requires knowledge of the entire probability distributions, which for distributions of drug properties among cell lines often turn out binomial, e.g. depending on corresponding gene. Hence for basic prediction mechanism there is proposed mixture of two Gaussians, trying to predict its weight based on additional information.
\end{abstract}
\section{Introduction}
In the age of precision medicine~\cite{prec}, there is a general trend of going from default therapies, toward personalized ones. It is especially important in cancer therapy, for which due to high mutation rate, there is extremely high variability between cell lines - bringing a difficult optimization question for personalized choice of chemotherapy.

There are many approaches for such individual optimizations based on additional information, e.g. tissue type, visual analysis of histopathology sample, genetic, transcriptonic, proteomic, etc. information. However, standard approaches (e.g. ~\cite{pred1,pred2,pred3}) are focused on prediction of value of specific properties (e.g. IC50, AUC) - strongly simplified description of complex behavior, properly described by probability distributions.

\begin{figure}[t!]
    \centering
        \includegraphics[width=85mm]{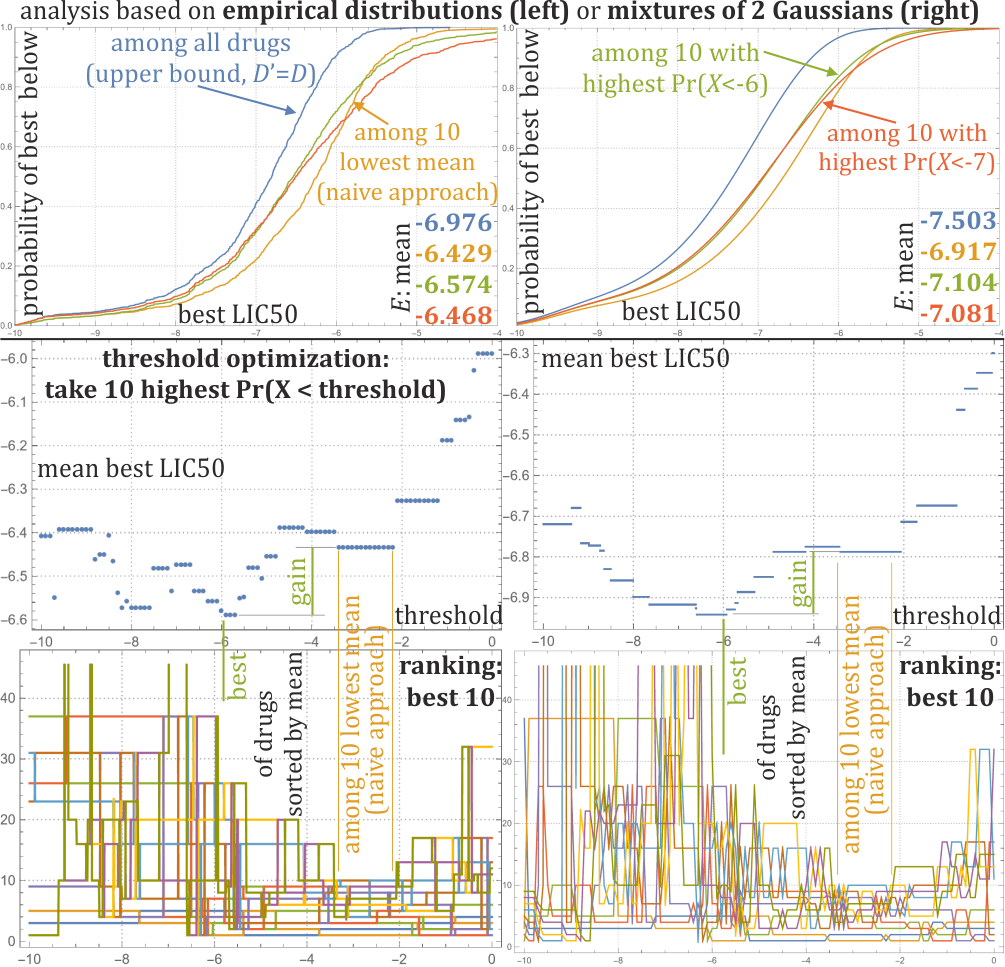}
        \caption{Superiority of probability distribution modelling for selection of the most promising drugs, analyzed and evaluated in two ways: using empirical distributions (left, slightly worse due to missing values) or parametric as mixtures of two Gaussians (right). Top: for an unknown cell line, probability of finding drug with LIC50 (natural logarithm of fitted IC50) below given value - among all 537 drugs (blue), or among 10 chosen in various ways: having the lowest mean (orange, naive approach), or proposed with quantile ranking: with the highest probability of being below given threshold - shown for -6 (green) and -7 (red). Middle: mean best LIC50 for various thresholds to optimize it ($\approx -6$). Bottom: visualization of rankings for various thresholds - positions of 10 best drugs, originally sorted by mean. We can see that threshold $\approx -3$ would give the naive approach, while the optimal one is $\approx -6$, bringing $\approx 0.15$ improvement for mean best LIC50. Examples of such selected drug batches are shown in Fig. \ref{dist}, \ref{thres}. }
       \label{intr}
\end{figure}

This article shows superiority of working on entire probability distributions of such values (for GDSC dataset~\cite{GDSC}), proposing and testing basic tools for this purpose. Basic advantage can be seen it top diagram of Fig. \ref{intr}: the best possible drug among 537 has distribution shown in blue. Orange line corresponds to testing 10 drugs chosen based only on (e.g. predicted) expected value, while for the green line there was optimized choice of 10 drugs using the entire probability distributions, getting us much closer to the best drugs for a given case. Figure \ref{thres} shows improvements for $n=1,\ldots,20$ size batches.

\begin{figure*}[t!]
    \centering
        \includegraphics[width=180mm]{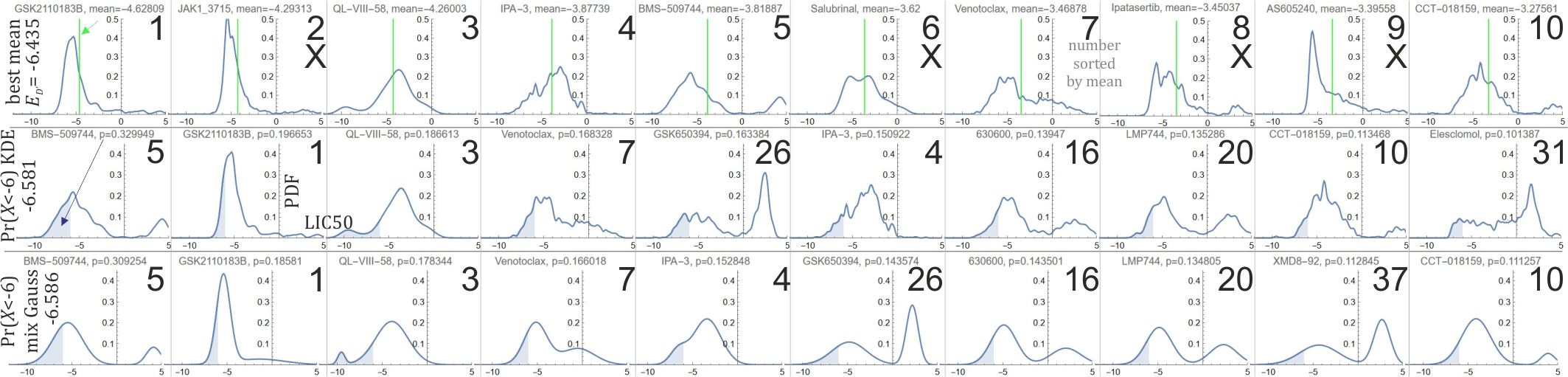}
        \caption{Chosen in 3 ways $n=10$ size drug batches for the first test of unknown cell line (no additional information, assuming it agrees with dataset statistics). Top: naive choice as those with the lowest mean for the minimised feature (LIC50). Lower two lines were chosen based on the proposed quantile ranking instead: maximizing probability of being below optimized threshold, -6 here. It allowed to improve $E_{D'}$ as expected value for the best drug in such batch by $\approx 0.15$. Large numbers show position if sorted by mean, "X" marks removed drugs. The top two lines show nearly accurate PDFs among cell lines (from kernel density estimation). In contrast, the bottom line models distribution as mixture of two Gaussians - it is less accurate, only slightly modifying ranking. Its main purpose is further prediction of probability distribution: based on additional information and/or previous test results, there is predicted weight between the two Gaussians in mixture. }
       \label{dist}
\end{figure*}

\begin{figure}[t!]
    \centering
        \includegraphics[width=85mm]{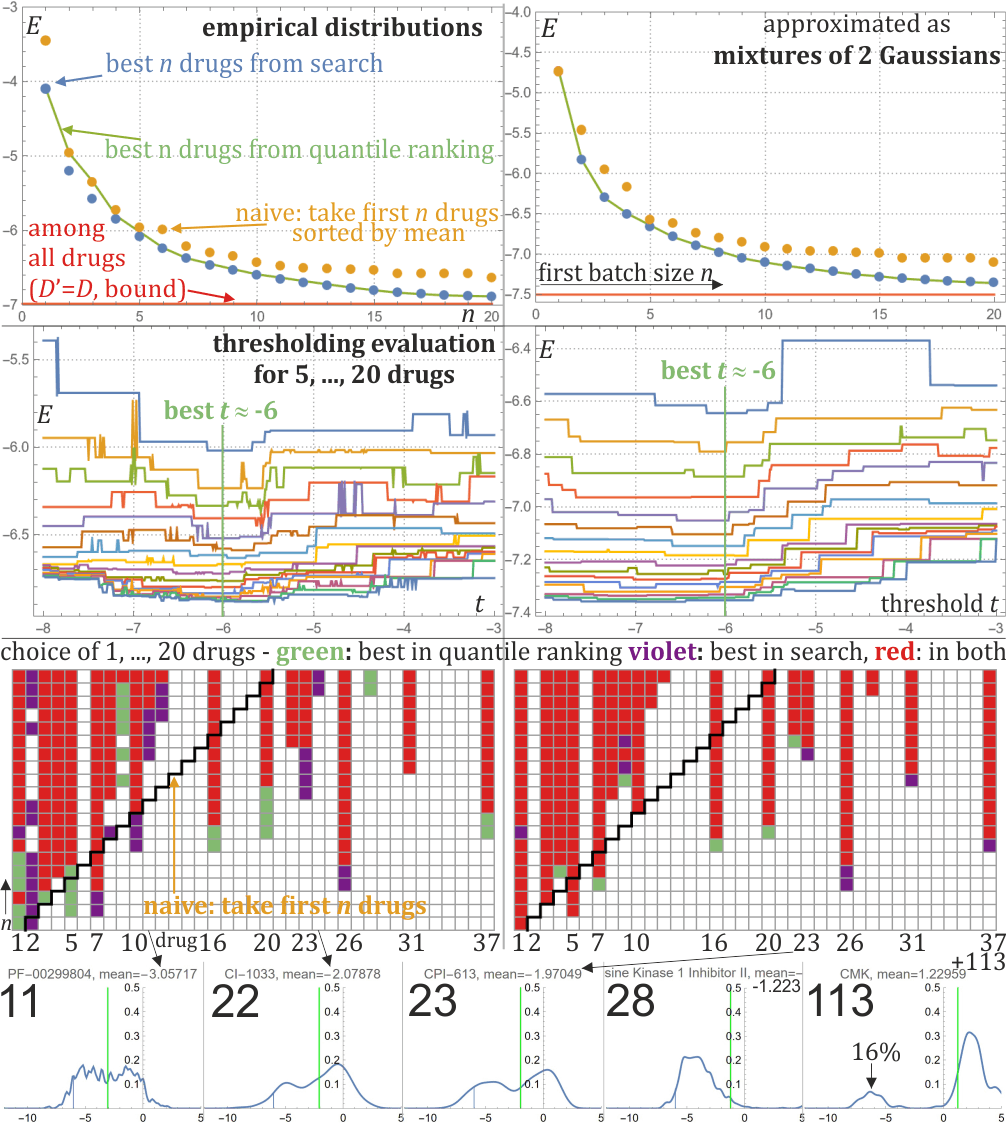}
        \caption{For empirical distribution (left) and its Gaussian mixture approximation (right), there were tested discussed approaches. Top: evaluation of mean best value for all drugs (red), and size $n=1,\ldots,20$ batches for various approaches. We can see naive approach (orange) using mean (e.g. predicted) is suboptimal and has much slower convergence to bound (red). Inexpensive thresholding optimization (green) is usually nearly as good as expensive search (blue). Middle: evaluating quantile ranking for $n=5,\ldots,20$ size batches, we can see universality - just taking $t=-6$ here usually gives nearly the best evaluation. Bottom: the best found drugs for $n=1,\ldots,20$ batch size, their comparison between quantile ranking and search approaches. We can see the differences are relatively small. There appear further drugs worth to include in considerations - some shown below and in Fig. \ref{dist}, e.g. 113-th drug accordingly to mean turns out promising in $\approx 16\%$ of cases.  }
       \label{thres}
\end{figure}

The main reason allowing for such significant improvements are often strongly binomial distributions of drug properties among cell lines (examples in Fig. \ref{dist}, \ref{thres}), corresponding e.g. to state of corresponding gene of the cells. Prediction of value as estimate of e.g. expected value is not expressive enough, especially for binomial distribution, and extreme statistics we are focused on - optimization of batch selection not for its mean, but for the best drug inside.

There are proposed basic tools to optimize the batch selection of drugs directly from data, and from parametric distributions - with prediction of their parameters based on additional information e.g. from results of test of the previous batch, or available additional information. We discuss simple mechanism of prediction of weight between two Gaussians in their mixture, in future to be compared with more complex probability prediction methods like Hierarchical Correlation Reconstruction (\cite{hcr1,hcr2}) decomposing statistical dependencies.

\section{Dataset and its initial processing}
While the presented approaches are very general, the shown diagrams were calculated from GDSC dataset~\cite{GDSC}: downloaded\footnote{https://www.cancerrxgene.org/} GDSC1 and GDSC2 datasets were merged, there were removed drugs and cell lines having available less than 1/4 of values, leading to $d=537$ drugs and $l=962$ cell lines. Some values were measured multiple times (up to 4) - in which case there is  used their average, however, in future their distributions can be included to improve discussed modeling of probability distributions.

This way we build $d\times l$ matrix $M$ containing the values of interest - here we focus on LIC50 as natural logarithm of fitted IC50 (half-maximal inhibitory concentration): the lower it is, the lower concentration of this drug is sufficient. Alternatively there could be used AUC (area under curve) evaluation, or maybe it is worth to extend to some more sophisticated evaluation techniques, include side effects, etc.

This $M$ $537\times 962$ matrix is missing 11\% of values. If possible, we use only the available values, for example to estimate distributions of values for given drug among cell lines. To evaluate chosen drugs and to build prediction models, currently there was used a basic cautious approach for imputation of the missing data. Specifically, missing $M_{ij}$ value is imputed as mean of two averages: for available $i$-t drug, and $j$-th cell line. As we are focused on searching for the best drugs, such cautious imputation leads to relatively uninteresting values - hence empirical distribution evaluation in Fig. \ref{intr} is slightly worse than based on mixed Gaussians. More sophisticated imputation techniques might be worth to consider in the future.

\section{Extreme subset optimization}
For an unknown cell line e.g. from histopathology sample, the task is to find one the best drugs among available, for example having the lowest LIC50. For this purpose we would like to test a selected batch of drugs, maybe followed by one or more such succeeding tests. For simplicity we assume that each such batch contains $n$ drugs, chosen e.g. as $n=10$. 

Our question of focus is: how to choose such $n$ drugs for the tests? Now from fixed distributions, generally for distributions predicted based on additional information.

As finally there is usually applied a single best drug, we will focus on its optimization: choose size $n$ subset of drugs $D'\subset D$ with the highest mean for the best drug inside such subset: $E_{D'}$. Therefore, while naively we would just choose drugs with the highest expected values (e.g. as results of value predictions), wanting to optimize such extreme statistics: the best in subset, we need to include into considerations the entire probability distributions, leading to essential improvements.

\subsection{Basic optimization criterion}
For the set of considered drugs $D=\{1,\ldots,d\}$, assume we have some models of probability distribution for each drug $X_i$, given by CDF (cumulative distribution function) $C_i:\mathbb{R} \to [0,1], C_i (x)=\textrm{Pr}(X_i\leq x)$ for drug $i\in D$.

For drug subset $D'\subset D$, assuming independence, probability of not containing drug below some threshold is one minus product of probabilities for all of them being above this threshold:
\be C_{D'}(x) =\textrm{Pr}\left(\left(\min_{i\in D'} X_i\right) \leq x\right)= 1- \Pi_{i\in D'} (1-C_i(x)) \ee
It allows to calculate the expected value, here of variable for the best drug in subset: $E_{D'}=\int_x x C'_{D'}(x) dx$, where $C'=\partial C/\partial x$ is derivative giving PDF (probability distribution function), eventually discretized if needed.

Finally the problem we focus on is finding size $n$ subset $D'$ minimizing expected value for the best drug there:
\be \textrm{argmin}_{D'\subset D, |D'|=n}  E_{D'} = \textrm{argmin}_{D'} \int_x x\, C'_{D'}(x) dx \label{opt}\ee
A naive choice of $D'$ subset is taking $n$ drugs with the lowest mean value in dataset - it is treated as the baseline approach we want to improve from.

\subsection{Quantile ranking optimization and growing search}
While optimization of (\ref{opt}) seems a difficult problem, we focus here on a simple inexpensive, but looking promising approximation: choosing a threshold $t$ and taking $n$ drugs with the highest probability of being below this threshold: \be \textrm{batch selection: take }n\textrm{ drugs having the highest }C_i(t) \ee
There is a nontrivial question of choosing this threshold $t$ - there was tested a discrete lattice for this parameter, and chosen the one leading to the lowest expected value $E_{D'}$, which usually turns out $\approx -6$ here.

Tested alternative was growing search, as taking the best found (e.g. 100 here) for size $n$ batches and adding all possible single drugs there, then taking e.g. 100 best ones - getting $n\to n+1$ step, to be started with $n=0$ empty set.

Such search is computationally much more expensive, in Fig. \ref{thres} results of various approaches are compared - we can see that inexpensive quantile ranking for $t\approx -6$ is usually nearly as good.

The above is sorting by "which quantile is given position $t$", alternatively we could sort by value of some specific quantile $C_i^{-1}(q)$, optimizing such $q$ instead of $t$. Additionally, it would allow to use quantile regression~\cite{quantile} techniques to directly predict them, to be tested in the future.

\subsection{Evaluation directly from matrix $M$, empirical distribution }
Having the GDSC  $d\times l$ matrix $M$ of measured values, we can directly estimate mean value of the best drugs from $D'\subset D$ subset:
\be E_{D'} = \frac{1}{l} \sum_{j=1}^l \min \{M_{ij}: i\in D'\}\label{ee}\ee
If instead of averaging in (\ref{ee}), we would sort the $(\min \{M_{ij}: i\in D'\})_j$ values, we get empirical distribution function estimation of CDF,  presented in top left diagram of Fig. \ref{intr}.

Using $D'=D$ all drugs we get the best possible performance for this dataset (blue plot in Fig. \ref{intr}), we search for let say $|D'|=10$ size subset minimizing $E_{D'}$. Naive approach is choosing drugs with the lowest mean, here estimated among available values (orange plot).

For quantile ranking a natural approach is using empirical distribution. We calculate $\textrm{Pr}(X_i\leq t)$ as percentage of available values for $i$-th drug being below threshold $t$. For various thresholds $t$, as $D'$ there were found $n=10$ drugs with the highest $\textrm{Pr}(X_i\leq t)$ (green, red plot).

As in the middle row of Fig. \ref{intr} we can choose threshold leading to $D'$ having the lowest $E_{D'}$, here $\approx -6$. Such evaluation plot is discrete as evaluating discrete selection of $n=10$ best drugs. This Figure also contains rankings for various thresholds, among drugs numbered accordingly to the mean among cell lines - we can see that targeting threshold $\approx -3$, we get the naive approach.

Optimization for empirical distribution has advantages - simplicity, these are the real data, also including statistical dependencies between drugs. However, one issue are the missing values - which might contain the best drug for a given situation, but will not be included in such optimization - leading to a bit inferior evaluation (fortunately similar ranking). A larger problem is including additional information for prediction - what rather requires parametric distributions.

\subsection{Parametric distribution model - Gaussian mixture}
Wanting to include additional information in optimization of batch of drugs to be tested: for the first batch (e.g. tissue type, visual evaluation, genetic, proteomic, etc.), or results of tests from earlier batches if testing succeeding ones, we rather need a parametric probability distribution model - and try to predict some of its parameters from such additional information.

Here we use mixture model of two Gaussians (referred as $A$ and $B$), as it agrees well with behavior of many drugs - with choice of $A$ or $B$ behavior based on e.g. expression of corresponding gene. For $\rho_{N(\mu,\sigma)}$ PDF of Gaussian, for mixture of two we use PDF:
\be\rho(x) = w\, \rho_{N(\mu_A,\sigma_A)}(x) +(1-w) \rho_{N(\mu_B,\sigma_B)}(x) \ee
For each drug $i$, among available values for cell lines, there were estimated (maximum likelihood, using Wolfram Mathematica) such 5 parameters: two centers $\mu$ and standard deviations $\sigma$, and weight $w$. Examples are shown in Fig. \ref{dist}.

\subsection{Predicting weight for Gaussian mixture}\label{prw}
Imagining that choosing between the two Gaussians in mixture corresponds to some on/off switch for e.g. expression of characteristic gene, the weight $w$ describes percentage of population having this switch in position $A$ - here inside used dataset, might be worth modifying, predicting e.g. based on tissue type.

This on/off switch approximation suggests to try to fix Gaussian parameters ($\mu_A,\sigma_A, \mu_B,\sigma_B$), and only try to predict probability of position of this switch: $w$ weight.

Seeing a value $X=x$, let us estimate what is the probability that it comes from Gaussian $A$ (not $B$):
\be \textrm{Pr}(A|X=x) = \frac{\textrm{Pr}(A)\,\textrm{Pr}(X=x|A)}{\textrm{Pr}(X=x)}=\frac{w\,\rho_{N(\mu_A,\sigma_A)}(x)}{\rho(x)} \ee
Having estimated $\{\rho_i\}_{i=1..d}$ mixture Gaussian distributions for all drugs $i$, we can calculate $W_{ij}=\textrm{Pr}(A|X_i=M_{ij})\in [0,1]$ matrix estimating probability that given value comes from the first $(A)$ of two Gaussians - which is kind of the best weight for this situation.

Generally not knowing this $W_{ij}$ local preferred weight, we would like to predict it from additional information, like results of the previous test. Denote  $c_j$ as context: vector of such additional information for $j$-th cell line, e.g. its results of earlier measurements, genetic features, etc.

\begin{figure}[t!]
    \centering
        \includegraphics[width=85mm]{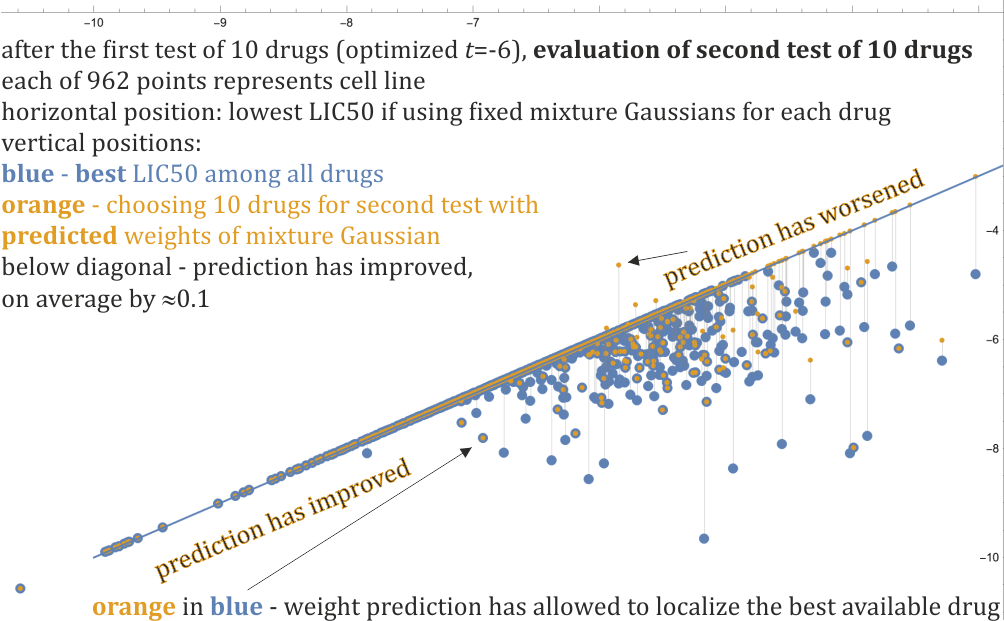}
        \caption{Evaluation of prediction of probability distribution to choose 10 drugs for the second test, based on results of the first test. Orange values above diagonal mean worsening, below mean improvement. Blue values show the best possible drugs for this line. As described in Section \ref{prw}, there were first calculated $W_{ij}$ probabilities of being in the first out of two Gaussians in mixture model, prediction model tries to predict $W_{ij}$ based on results from 10 drugs in the first test (using (\ref{prm})). For each such predicted distribution, there was used threshold $t$ minimizing $E_{D'}$.        }
       \label{second}
\end{figure}

Here we perform such prediction of $W_{ij}$ with least-squares linear regression from contexts $(c_j)_j$, independently for all drugs of interest ($i$):
\be \textrm{argmin}_{\beta^i} \sum_j \left(c_j\cdot \beta^i- f(W_{ij}) \right)^2 \label{prm}\ee
\begin{itemize}
  \item To predict value from $\mathbb{R}$ (while $W_{ij}\in [0,1]$), there is applied $f$ function here being inverse CDF of normalized Gaussian ($\mu=0,\sigma=1$), hence the prediction result is finally transformed with its CDF,
  \item Context $c_j$ contains '1' to include intercept term of linear regression,
  \item As providing better performance, used context $c_j$ contains both values from $M_{ij}$ of previous test, and also of $f(W_{ij})$ matrix to include convenient nonlinearity,
  \item As providing better performance, there were finally used $(W'_{ij}+w_i)/2$ more cautious weights for mixed Gaussian, where $W'_{ij}$ is predicted $W_{ij}$ with above linear regression, and $w_i$ is the original weight for $i$-th drug.
\end{itemize}
Example of evaluation of such procedure is shown in Fig. \ref{second} - of choice of the second batch, based on results from the first batch, allowing for  improvement.

Analogously there can be included other information, e.g. available before the first test. Linear regression can be replaced with a more sophisticated methods like neural networks, there can be added regularization, or considered more sophisticated probability prediction approaches like Hierarchical Correlation Reconstruction - planned for further work.

\section{Conclusions and further work}
There were briefly presented advantages and basic methodology for considering, predicting entire probability distribution for selection of the most promising drugs to test.

This is early version of article with intention to present this looking promising novel way of thinking, with many perspectives for planned further work, for example:

\begin{itemize}
  \item Exploit additional information for prediction of probability distributions, to choose especially the drugs for the first test batch.
  \item Use different basic distributions, e.g. mixture of distributions from a more general family like exponential power distribution ($\rho\sim \exp(-|x|^\kappa)$), or heavy tailed like stable, student t distribution, maybe mixture of more than 2 distributions.
  \item Use different probability prediction methods, like neural networks instead of linear regression for weights, or more sophisticated methods like quantile regression~\cite{quantile}, Hierarchical Correlation Reconstruction (HCR)~\cite{hcr1,hcr2}.
  \item There was assumed independence inside optimized batch ((\ref{opt}) formula) - it might be worth to include their statistical dependence, possible e.g. with HCR.
  \item Consider more appropriate drug evaluations than IC50 e.g. AUC, maybe combining them, include side effects, etc.
  \item Include details of testing for estimation of e.g. IC50 into optimization, e.g. increasing the number of tested drugs at cost of reduced number of tested concentrations.
  \item We focus on finding the lowest IC50 drug, while in some scenarios it might be worth finding a few of them, or applying some additional criteria, restrictions, evaluation factors.
\end{itemize}

\bibliographystyle{IEEEtran}
\bibliography{cites}
\end{document}